%% file: main.tex
\documentclass[sigconf]{acmart}
\input{predefine}
\usepackage[utf8]{inputenc}
\usepackage[most]{tcolorbox}
\usepackage{listings}
\usepackage{upquote}
\usepackage{subfigure}
\usepackage{enumitem}
\setlist{leftmargin=2mm}
\usepackage[font=footnotesize]{caption}
\usepackage{xcolor}
\usepackage{soul}
\soulregister{\cite}7
\soulregister{\citep}7
\soulregister{\citet}7
\soulregister{\ref}7 
\soulregister{\pageref}7
\soulregister{\tool}{7}
\soulregister{\item}7
\soulregister{\subsection}7

\usepackage{algorithm}
\usepackage{algpseudocode}
\usepackage{longtable}

\usepackage{CJK}

\lstdefinestyle{JAVA}{
  language=JAVA,
  moredelim=[is][\underbar]{_}{_},
}

\newcommand{\tool}{JUnitTestGen}

\begin{document}

\title{
Mining Android API Usage to Generate Unit Test Cases for Pinpointing Compatibility Issues
}

\author{Xiaoyu Sun}
\email{xiaoyu.sun@monash.edu}
\affiliation{
  \institution{Monash University, Australia}
  \streetaddress{Wellington Rd}
  \city{Clayton}
  \state{VIC}
  \postcode{3800}
}

\author{Xiao Chen}
\email{Xiao.chen@monash.edu}
\affiliation{
  \institution{Monash University, Australia}
  \streetaddress{Wellington Rd}
  \city{Clayton}
  \state{VIC}
  \postcode{3800}
}

\author{Yanjie Zhao}
\email{Yanjie.Zhao@monash.edu}
\affiliation{
  \institution{Monash University, Australia}
  \streetaddress{Wellington Rd}
  \city{Clayton}
  \state{VIC}
  \postcode{3800}
}

\author{Pei Liu}
\email{Pei.Liu@monash.edu}
\affiliation{
  \institution{Monash University, Australia}
  \streetaddress{Wellington Rd}
  \city{Clayton}
  \state{VIC}
  \postcode{3800}
}

\author{John Grundy}
\email{john.grundy@monash.edu}
\affiliation{
  \institution{Monash University, Australia}
  \streetaddress{Wellington Rd}
  \city{Clayton}
  \state{VIC}
  \postcode{3800}
}

\author{Li Li}
\authornote{Li Li is the corresponding author.}
\email{li.li@monash.edu}
\affiliation{
  \institution{Monash University, Australia}
  \streetaddress{Wellington Rd}
  \city{Clayton}
  \state{VIC}
  \postcode{3800}
}

\begin{abstract}
Despite being one of the largest and most popular projects, the official Android framework has only provided test cases for less than 30\% of its APIs.
Such a poor test case coverage rate has led to many compatibility issues that can cause apps to crash at runtime on specific Android devices, resulting in poor user experiences for both apps and the Android ecosystem.
To mitigate this impact, various approaches have been proposed to automatically detect such compatibility issues.
Unfortunately, these approaches have only focused on detecting signature-induced compatibility issues (i.e., a certain API does not exist in certain Android versions), leaving other equally important types of compatibility issues unresolved. 
In this work, we propose a novel prototype tool, \tool{}, to fill this gap by mining existing Android API usage to generate unit test cases.
After locating Android API usage in given real-world Android apps, \tool{} performs inter-procedural backward data-flow analysis to generate a minimal executable code snippet (i.e., test case).
Experimental results on thousands of real-world Android apps show that \tool{} is effective in generating valid unit test cases for Android APIs. We show that these generated test cases are indeed helpful for pinpointing compatibility issues, including ones involving semantic code changes.

\end{abstract}
\maketitle
\section{Introduction}

Unit testing is a type of software testing aiming at testing the effectiveness of a software's units, such as functions or methods.
It is often the first level of testing conducted by software developers themselves (hence a white box testing technique) to ensure that their code is correctly implemented.
Unit testing has many advantages.
First, it helps developers fix bugs early in the development cycle which subsequently saves costs in the end. Indeed, it is known that the cost of a bug increases exponentially with time in the software development workflow.
Second, it makes it possible to achieve regression testing. For example, if developers refactor their code later, it allows them to make sure that the refactored code still works correctly.
Third, unit tests provide an effective means for helping developers understand the unit under testing, i.e., what functions are provided by the unit and how to use them?
Because of the aforementioned advantages, it is recommended to always write unit tests when developing software, and the unit tests should cover as many units as possible.

The Android framework, as one of the largest software projects (with over 500,000 commits), is no exception. 
The Android framework provides thousands of public APIs that are heavily leveraged by app developers to facilitate their development of Android apps. Ideally, each such public API should be provided with a set of unit tests to ensure that the API is correctly implemented and the continuous evolution of the framework will not change its semantics.
Unfortunately, based on our preliminary investigation, less than 30\% of APIs, are provided with unit test cases, leaving the majority of APIs uncovered.
This is unacceptable considering that the Android framework nowadays has become one of the most popular projects (with millions of devices running it).

This poor test coverage of Android APIs has led to serious compatibility issues in the Android ecosystem, as recently shown~\cite{li2018cid,wei2016taming,mutchler2016target,zhang2015compatibility,ham2011mobile,huang2018understanding}.
For example, Li et al. ~\cite{li2018cid} demonstrate that various Android APIs suffer from compatibility issues as the evolution of the Android framework will regularly remove APIs from or add APIs into the framework. Such API removal or addition can result in no such class or method runtime exceptions when the corresponding app is running on certain framework versions.
Liu et al.~\cite{liu2021} further present an approach to detect silently-evolved Android APIs, which could cause another type of compatibility issue as their semantics are altered (while not explicitly documented) due to the evolution of the Android framework.
Moreover, Wei et al.~\cite{wei2019pivot} experimentally show that some Android APIs could even be customized by smartphone manufacturers, leading to another type of compatibility issue that causes Android apps to crash on certain devices while behaving normally on others.
The authors further propose a prototype tool called PIVOT to automatically learn device-specific compatibility issues from existing Android apps. Their experiments on a set of top-ranked Google Play apps have discovered 17 device-specific compatibility issues.

To the best of our knowledge, the state-of-the-art works targeting compatibility issue detection leverage static analysis techniques to achieve their purpose.
However, as known to the community, the static analysis will likely yield false positive results as it has to make some trade-offs when handling complicated cases (e.g., object-sensitive vs. object-insensitive).
In addition, the static analysis will also likely suffer from soundness issues because some complicated features (e.g., reflection, obfuscation, and hardening) are difficult to be handled~\cite{sun2021taming,samhi2022jucify}.
Furthermore, except for syntactic changes, compatibility issues could also be triggered by semantic changes, which are non-trivial to be handled statically.
Indeed, as demonstrated by Liu et al.~\cite{liu2021}, there are various semantic change-induced compatibility issues in the Android ecosystem that remain undetected after various static compatibility issue detection approaches are proposed to the community.

Moreover, static app analysis can only be leveraged to perform post-momentum analysis (i.e., after the compatibility issues are introduced to the community). 
They cannot stop the problems from being distributed into the community -- many Android apps, including very popular ones, still suffer from compatibility issues.
To mitigate this problem, we argue that incompatible Android APIs should be addressed as early as possible, i.e., ideally, at the time when they are introduced to the framework.
This could be achieved by providing unit tests for every API introduced to the framework and regressively testing the APIs against Android devices with different manufacturers and different framework versions.
However, it is time-consuming to manually write and maintain unit tests for each Android API (which probably explains why there is only a small set of APIs covered by unit tests at the moment).
There is hence a need to automatically generate compatibility unit tests for Android APIs.


In this work, we present a prototype tool, \tool{}, that attempts to automatically generate test cases for Android APIs based on their practical usage in real-world apps.
Specifically, after locating existing API usages in real-world Android apps, \tool{} performs field-aware, inter-procedural backward data-flow analysis to infer the API caller instance and its parameter values.
\tool{} then leverages the inferred information to reconstruct a minimal executable code snippet for the API under testing.
Experimental results on thousands of Android apps show that \tool{} is effective in generating test cases for Android APIs. It achieves an 80.4\% of success rate in generating valid test cases. 
These test cases subsequently allow our approach to pinpoint various types of compatibility issues, outperforming a state-of-the-art generic test generation tool named EvoSuite, which can only generate test cases to reveal a small subset of compatibility issues.
Furthermore, we demonstrate the usefulness of \tool{} by comparing it against a state-of-the-art static analysis-based compatibility issue detector called CiD. \tool{} is able to mitigate CiD's false-positive results and go beyond CiD's capability (i.e., detecting compatibility issues induced by APIs' signature changes) to detect compatibility issues induced by APIs' semantic changes.

Overall, we make the following main contributions in this work:
\begin{itemize}
\item We have designed and implemented a prototype tool \tool{} that leverages a novel approach to automatically generate unit test cases for APIs based on their existing usages.

\item We have set up a reusable testing framework for pinpointing API-induced compatibility issues by automatically executing a large set of unit test cases on multiple Android devices. 

\item We have demonstrated the effectiveness of \tool{} by i) generating valid test cases for Android APIs and pinpointing problematic APIs that could induce compatibility issues if accessed by Android apps, ii) outperforming state-of-the-art tools on real-world apps in detecting a wider range of compatibility issues.
\end{itemize}

The source code\footnote{\url{https://github.com/SMAT-Lab/JUnitTestGen}} and experimental results are all made publicly available in our artifact package.\footnote{\url{https://doi.org/10.5281/zenodo.6507579}}.

\section{Motivation}

To overcome the fragmentation problem, our fellow researchers have proposed various approaches to mitigate the usage of compatibility issues in Android apps~\cite{wei2016taming, wei2019pivot, xia2020android}.
These approaches mainly leverage static analysis to achieve their purpose.
Unfortunately, static analysis is known to likely generate false-positive and false-negative results and is yet hard to handle such issues that involve semantics changes in Android APIs~\cite{liu2021identifying}.
Therefore, we argue that there is a need also to invent dynamic testing approaches to complement existing works in handling app compatibility issues.

We hence start by conducting a preliminary study investigating the test case coverage in the Android framework. Specifically, we downloaded the source code of AOSP from API level 21 to 30 and then calculated the number of public APIs\footnote{The APIs in \texttt{platform/frameworks/base} path.} and their corresponding unit test cases provided by Google. Our result reveals that on average \textbf{less than 30\% of Android framework APIs have provided test cases in each API level}, indicating the Android framework has a poor test case coverage.
When more APIs are provided with unit test cases, more compatibility issues of APIs will likely be identified during regression testing. This will enable them to be fixed at an earlier stage to avoid the introduction of compatibility issues in the first place. 
To this end, we propose to effectively and efficiently detect compatibility issues through a dynamic testing approach that fulfills its objective by automatically generating valid test cases by mining API usages from real-world Android apps.

\begin{figure*}[!h]
    \centering
    \setlength{\abovecaptionskip}{0.2pt}
    \setlength{\belowcaptionskip}{0.2pt}
    \includegraphics[width=0.9\textwidth]{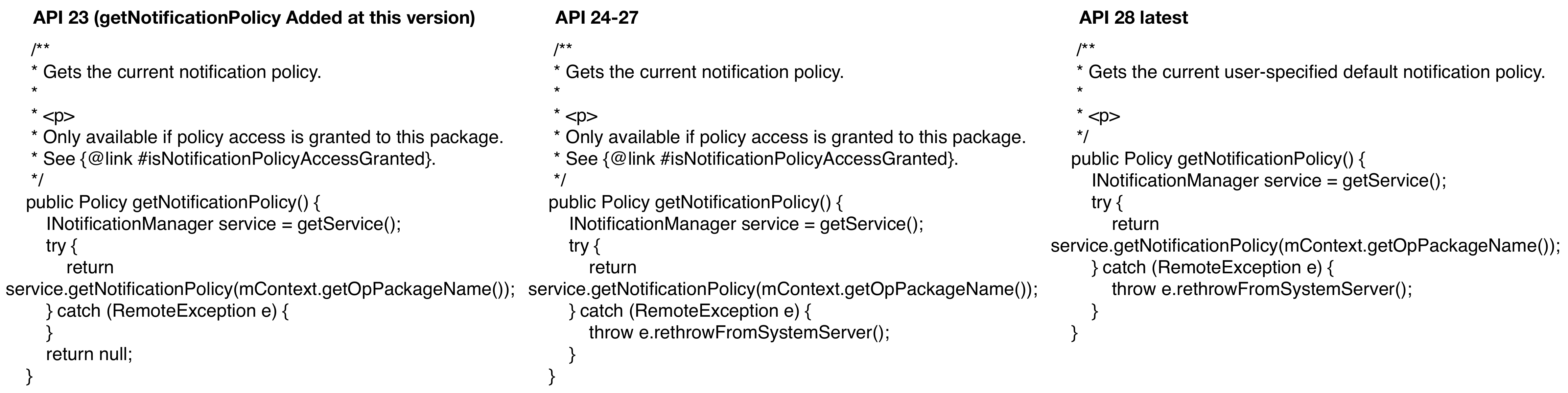}
	\caption{The source code of Android API \emph{getNotificationPolicy()}.}
    \label{fig:getNotificationPolicy}
\end{figure*}


\textbf{Why Dynamic Testing.} We now present a concrete example to motivate why there is a need to generate more and better unit test cases for Android APIs to pinpoint compatibility issues.
There is an Android API called \emph{getNotificationPolicy()}, located in class \emph{NotificationManager}.
At the moment, there are no unit tests provided for this API.
The lack of solid testing for this API has unfortunately led to various problems, as demonstrated by the various discussions on StackOverflow~\cite{stackoverflowPermissions}, one of the most widely used question and answer websites.

Figure~\ref{fig:getNotificationPolicy} presents the brief evolution of the source code of \emph{getNotificationPolicy()}.
This API was introduced to the Android framework at API level 23.
The apps that accessed this API would crash on devices powered by Android with API level 22 or earlier, resulting in forwarding compatibility issues.
During the evolution of the Android framework, the implementation of \emph{getNotificationPolicy()} is quickly changed at API level 24 (i.e., no longer returns null).
Nevertheless, at this point, the comments are not changed, suggesting potential compatibility issues because of changed semantics.
At API level 28, this API is further changed.
This time, only the comments are changed, i.e., the implementation of the API is kept the same.
This change further suggests that this API may be involved with compatibility issues as the source code of the API and its comments are inconsistent at a certain period (over six API levels).

\begin{lstlisting}[
caption={Code example of invoking API \emph{getNotificationPolicy()}.},
label=code:getNotificationPolicy_app, firstnumber=1, abovecaptionskip=0pt, belowcaptionskip=0pt, aboveskip=0.5pt, belowskip=0.8pt]
@Override
protected void onCreate(Bundle savedInstanceState) {
 super.onCreate(savedInstanceState);
 setContentView(R.layout.activity_main);
 
 NotificationManager mng = (NotificationManager) this.getSystemService("notification");
 NotificationManager.Policy policy = mng.getNotificationPolicy();
 int priorityCallSenders = policy.priorityCallSenders;
}\end{lstlisting}

The actual implementation of \emph{getNotificationPolicy()} is through the complicated inter-process communication mechanism (e.g., the API is defined via Android Interface Definition Language (AIDL). Thus, it is non-trivial to confirm if there is a compatibility issue after API level 23 by only (statically) looking at the Java code of the framework. It is still a known challenge to statically analyze Java and C code at the same time.

We resort to a dynamic approach to check if \emph{getNotificationPolicy()} suffers from compatibility issues.
Specifically, we implement a simple Android app with minimal lines of code to invoke the API (i.e., lines 6-7 in Listing~\ref{code:getNotificationPolicy_app}) and also the usage of the return value of this API(i.e., lines 8 in Listing~\ref{code:getNotificationPolicy_app}).
The minimal and targeted SDKs of this app are set to be 21 and 30, respectively.
We then launch this app on ten emulators covering Android API levels from 21 to 30.
As expected, the app throws \emph{NoSuchMethodError} as the API is not yet introduced at API level 21 and 22.
From API level 23 to 27, the app throws \emph{SecurityException} with the message ``Notification policy access denied'', due to a lack of declaration of Android permissions. Even though the permission was granted, the API can still introduce compatibility issues at API level 23 because the return value can be null, causing \emph{NullPointerException} later on (e.g., lines 8 in Listing 1)). 
Surprisingly, since API level 28, the app does not throw any exception, even though no permissions are declared as well.
We then go one step further to track the detailed implementation of \emph{service.getNotificationPolicy(String)} and found that the enforcement of policy access (i.e., line 3 in Listing~\ref{code:getNotificationPolicy_change}) is removed in the API at API level 28, which explains why the app no longer crashes since API level 28.

\begin{lstlisting}[
caption={Code changes of method \emph{getNotificationPolicy(String)}, which includes the underline implementation of Android API \emph{getNotificationPolicy()}.},
label=code:getNotificationPolicy_change,
firstnumber=1,abovecaptionskip=0pt, belowcaptionskip=0pt, aboveskip=0.5pt, belowskip=0.5pt]
 @Override
 public Policy getNotificationPolicy(String pkg) {
-  enforcePolicyAccess(pkg, "getNotificationPolicy");
  final long identity = Binder.clearCallingIdentity();
  try {
   return mZenModeHelper.getNotificationPolicy();
  } finally {
   Binder.restoreCallingIdentity(identity);
 }}
\end{lstlisting}

These observed runtime behaviours strongly indicate that the direct invocation of \emph{getNotificationPolicy()} will very likely result in two compatibility issues: (1) a method is not yet defined and (2) method semantics have been altered.
Ideally, such issues -- especially the latter case -- should not be introduced into the Android framework.
However, due to the lack of API unit tests, such issues are non-trivial to identify and avoid.
It is hence essential to provide more and better unit test cases for all Android APIs.
Since it is time-consuming to achieve this manually, we argue that there is a strong need to provide automated approaches to automatically generate unit test cases for Android APIs to identify compatibility issues as early as possible.
In this work, we propose to generate unit test cases for Android APIs by learning from existing Android API usages.

\textbf{Why mining API usage.} In addition, generic test generation tools~\cite{fraser2011evolutionary, pacheco2007randoop,Kex} mainly targeting on satisfying coverage criteria for classes, while compatibility issues are mainly caused by the fast-evolving of APIs~\cite{li2018cid}, which makes them insufficient in detecting compatibility issues. Specifically, such generic test case generation approaches (such as EvoSuite) are tailored to generate tests based on the source code of classes, lacking API usage knowledge~\cite{kechagia2019effective}, including both API calling context and API dependency knowledge. This information is crucial to setting up the environment for successfully calling Android APIs properly to detect compatibility issues. In addition, all of these tools completely ignore semantic-level behaviours at the API level, leading to many compatibility issues undetected. Especially in Android, APIs often come with usage
caveats, such as constraints on call order~\cite{ren2020api}. Thus, it is essential to capture API dependencies~\cite{zhang2014semantics} involved in the calling context before invoking the target API. However, it is challenging for traditional test generation tools to achieve this since they generate test suites only based on source code, which lacks API dependency knowledge. Thus, the insufficiency of the generic coverage-based test case generation approach motivates us to mine API usage to generate much more effective test cases in detecting compatibility issues.

\vspace{-10pt}

\section{Our Approach}
\begin{figure*}[!h]
    \centering
    \includegraphics[width=0.83\linewidth]{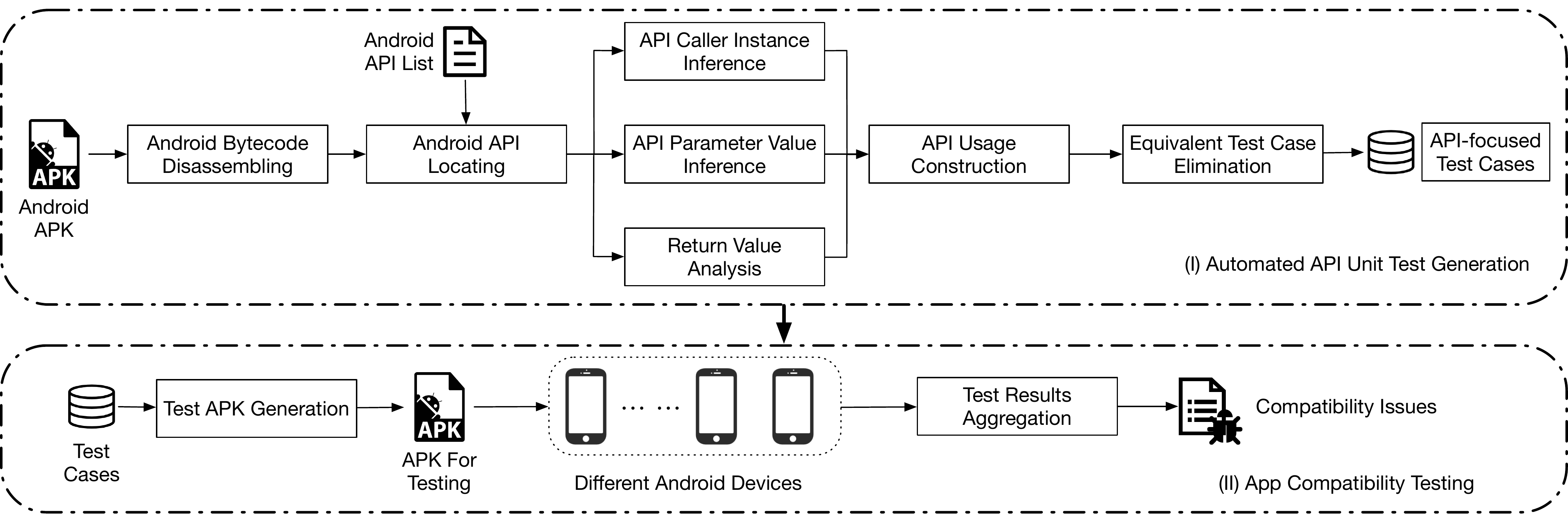}
	\caption{The working process of our approach.}
    \label{fig:methodology}
\end{figure*}

The main goal of this work is to automatically generate unit tests for the Android framework to provide better coverage at the unit test stage of as many Android APIs as possible. Fig.~\ref{fig:methodology} outlines the process of \tool{}, which is made up of two modules involving a total of nine steps.
We first locate target API invocations after disassembling the APK
bytecode. We then apply inter-procedural data-flow analysis to identify the API usage, including API caller instance inference and API parameter value inference. We then execute these generated test cases on Android devices with different Android versions (i.e., API levels). The following elaborates on the detailed process of each module.

The first module, \emph{Automated API Unit Test Generation}, includes five sub-processes to automatically generate unit test cases for Android APIs.
This module takes as input an Android API (or an API list) for which we want to generate unit tests and an Android app that invokes the API and outputs a minimum executable code snippet involving the given API.
This code snippet is the API's unit test case.

The second module, \emph{App Compatibility Testing}, includes two sub-processes.
This module takes as input the previously generated unit test cases to build an Android app, allowing direct executions of the test cases on Android devices running different framework versions (i.e., API levels). 
The output of this second module is the execution results of the test cases concerning different execution environments. Using this, \tool{} can then determine all the Android APIs suffering from potential compatibility issues.
We now detail these two modules of \tool{} below.

\vspace{-5pt}

\subsection{Automated API Unit Test Generation}
\label{sub:unit_test_generation}

As shown in Figure~\ref{fig:methodology}, the first module of \tool{}, namely Automated API Unit Test Generation, is made up of five steps.

\textbf{I.1 Android Bytecode Disassembling.} \tool{} takes Android APKs as input. 
Given an Android app, the first step of \tool{} is to disassemble the Dalvik bytecode to an intermediate representation. In this work, we leverage Soot~\cite{vallee2010soot}, which provides precise features for code analysis. In particular, it supports 3-address code intermediate  representation Jimple and accurate call-graph analysis framework Spark\cite{lhotak2003scaling}. On top of Soot, \tool{} is able to convert Android APK's bytecode into Jimple precisely.

\textbf{I.2 Android API Locating.}
The second step towards locating the practical usages of Android APIs is quite straightforward.
\tool{} visits every method of each application in its Jimple representation, i.e., statement by statement, to check if any of the APIs in the input list is invoked. If so, we record the location and mark the usage of the API as located.

\textbf{I.3 API Caller Instance Inference.}
Android APIs are invoked as either \emph{static methods} (also known as \emph{class methods}) or \emph{instance methods}. A static method can be called without needing an object of the class, while an instance method requires an object of its class to be created before it can be called.
Listing~\ref{code:CALL_static_instance_APIs} shows an example for each method type in the Jimple representation. The static method (\emph{boolean equals()}) can be called directly as long as the API signature (\emph{<android.text.TextUtils: boolean equals(java.lang.CharSequence, java.lang.CharSequence)>}) is acquired, which has already been done in step I.2. However, to invoke the instance method (\emph{void setLayoutParams()}), its calling object \emph{\$r5} needs to be identified. Specifically, a backward data-flow analysis is needed to locate both the definition statement of \emph{\$r5} and any intermediate APIs (or methods) on the call trace that may change the behaviour of the caller instance.

\begin{lstlisting}[
caption={Examples of calling static and instance APIs.},
label=code:CALL_static_instance_APIs,
firstnumber=1]
//Calling a Static API
<android.text.TextUtils: boolean equals(java.lang.CharSequence,java.lang.CharSequence)>

//Calling an Instance API
$r5.<android.widget.ListView: void setLayoutParams(android.view.ViewGroup$LayoutParams)>
\end{lstlisting}

Identifying the calling object and constructing the call trace is a non-trivial task. 
As the calling object's instantiation could depend on multiple method calls (e.g., can be defined in other methods and passed on via callee's returned values), the caller instance inference process needs to be inter-procedural.
Furthermore, each of the involved methods in the instantiation process may further require specific parameter values, which need to be properly prepared in order to successfully create the calling object.
Hence, the caller instance inference process requires backtracing not only the direct caller instance but also many other variables leveraged by the app to instantiate the calling object.
\setlength{\intextsep}{1.2pt} 
\begin{algorithm}
 \scriptsize
  \caption{API Caller Instance Inference.}\label{API_Caller_Instance_Inference}
  \begin{algorithmic}[1]
    \Require $api_t$: the target API
    \Ensure $M$: a list of method invocations towards $api_t$
    \Function {calculateMinimumExecutableContext}{$api_t$}
        \State $M = \emptyset$
        \State $stmt_t\gets $ statement contains $api_t$
        \If {$api_t$ is a static API}
            \State $M\gets$ The API signature of $api_t$
        \EndIf
        \If {$api_t$ is an instance API}
            \State $assignStmt\gets $ \Call{getDefinitionStmt}{$stmt_t$}
            \State $M\gets $ \Call{constructCallTrace}{$assignStmt, \emptyset$}
        \EndIf
        \State \Return{$M$}
    \EndFunction
  \end{algorithmic}
\end{algorithm}


\begin{algorithm}
\scriptsize
  \caption{Construct Call Trace.}\label{Construct_Call_Trace}
  \begin{algorithmic}[1]
    \Require $stmt_{co}$: A definition statement of the calling object
    \Statex $visitedStmts$ : The set of statements visited by the analysis.
    \Ensure $callTrace$: a list of method invocations towards $stmt_{co}$
    \Function{constructCallTrace}{$stmt_{co}, visitedStmts$}
        \State $callTrace \gets \emptyset$
        \If{($stmt_{co}$ in $visitedStmts$) \textsf{OR} ($stmt_{co}$ is Constant) \textsf{OR} ($stmt_{co}$ is Syetem API)}
            \State \Return{$callTrace$}
        \EndIf
        \State $visitedStmts \gets visitedStmts \cup stmt_{co}$
        \If{$stmt_{co}$ is ParameterRef}
            \State $stmt_{inv}\gets getInvolvingStmtsfromCallGraph(stmt_{co})$
            \For{$each\ s \in stmt_{inv}$}
                \State $definitionStmt_{param}\gets $ \Call{getDefinitionStmt}{$s$}
                \State $callTrace\gets $ \Call{constructCallTrace}{$definitionStmt_{param}, visitedStmts$}
            \EndFor
        \EndIf
        \If{$stmt_{co}$ is InvokeExpr}
            \State $stmt_{inv}\gets getInvolvingStmtsfromCallGraph(stmt_{co})$
            \For{$each\ s \in stmt_{inv}$}
                \State $method_{host} \gets $ the host method of $s$
                \State $returnStmt \gets getReturnStmtFromMethod(method_{host})$
                \State $definitionStmt_{returnStmt}\gets $ \Call{getDefinitionStmt}{$returnStmt$}
                \State $callTrace\gets $ \Call{constructCallTrace}{$definitionStmt_{returnStmt}, visitedStmts$}
            \EndFor
        \EndIf
    \EndFunction
  \end{algorithmic}
\end{algorithm}

Algorithm~\ref{API_Caller_Instance_Inference} gives the details of the approach. Given a target API as input, we apply a backward data flow analysis to identify the minimum executable context of the target API (lines 1\textasciitilde12). As shown in Algorithm~\ref{API_Caller_Instance_Inference}, we describe the API caller inference process for both static methods (lines 4\textasciitilde6) and instance methods (lines 7\textasciitilde10). For static methods, we return its corresponding API signature, which has been obtained in step I.2. For instance methods, we first locate the definition statement of the calling object through invoking method \emph{getDefinitionStmt} (line 8), which returns the definition statement for a local variable. This method walks the inter-procedural control flow graph from the target API statement in reverse order, aiming to look for the nearest assignment statement defining the API's calling object. 
After that, with the help of the function \emph{constructCallTrace}(i.e., defined in Algorithm ~\ref{Construct_Call_Trace}) (line 9), we can extract the call trace corresponding to the calling object. As shown in Algorithm~\ref{Construct_Call_Trace}, we handle parameter callers (lines 7\textasciitilde13) and method callers (lines 14\textasciitilde22), respectively. If the definition statement of the calling object comes from a parameter reference, we first retrieve all the statements at which the invocation occurs, and then for each of the statements, we recursively construct its call trace by calling the method \emph{constructCallTrace}. A similar process has been applied to handle method callers (lines 14\textasciitilde22), which recursively construct the call trace involving statements of the calling object. The recursive process will not terminate until any of the conditions have been satisfied in line 3, i.e., either the statement is a constant, or it has been visited before, or it is an Android system API.

We elaborate on this process with a Jimple code example presented in Listing~\ref{code:Code_Example_API_Caller_Instance_Return_Value}.
In this example, the target Android API to test (i.e., \emph{queryDetailsForUid(int,String,long,long,int)}) is invoked in line 26 by the calling object \emph{\$r3}, where \emph{\$r3} is a returned value of a self-defined method \emph{getNetworkStatsManager(Context)} (line 22). We then step into the definition of the method \emph{getNetworkStatsManager(Context)} (lines 1\textasciitilde7), and further backtrace the variables \emph{\$r2} and \emph{\$r1} along the call chain. \emph{\$r1} retrieves the network stats service from the application context \emph{\$r0} (line 4), and finally, the backtrace terminates at \emph{\$r0} (line 3), where all unknown variables are resolved. 

\begin{lstlisting}[
caption={Code example demonstrating the usage of API \emph{queryDetailsForUid}. The code snippet is extracted from app \emph{com.eyoung.myutils}.},
label=code:Code_Example_API_Caller_Instance_Return_Value,
firstnumber=1]
public static android.app.usage.NetworkStatsManager getNetworkStatsManager(android.content.Context)
{
 $r0 := @parameter0: android.content.Context;
 $r1 = $r0. getSystemService("netstats");
 $r2 = (android.app.usage.NetworkStatsManager) $r1;
 return $r2;
}

public static int getUid(android.content.Context) throws android.content.pm.PackageManager$NameNotFoundException
{
 $r0 := @parameter0: android.content.Context;
 $r1 = $r0.getPackageManager();
 $r2 = $r0.getPackageName();
 $r3 = $r1.getApplicationInfo($r2, 1);
 i0 = $r3.<android.content.pm.ApplicationInfo: int uid>;
 return i0;
}

public static float getCurAppFlow(android.content.Context) throws android.content.pm.PackageManager$NameNotFoundException
{
 $r0 := @parameter0: android.content.Context;
 $r3 = DeviceInfoUtil.getNetworkStatsManager($r0);
 $l1 = System.currentTimeMillis();
 $i0 = DeviceInfoUtil.getUid($r0);
 //Target API
 $r5 = virtualinvoke $r3.<android.app.usage.NetworkStatsManager: android.app.usage.NetworkStats queryDetailsForUid(int,java.lang.String,long,long,int)>(0, "", 0L, $l1, $i0);
}
\end{lstlisting}

In addition to inter-procedural data-flow analysis, \tool{} also needs to be field-aware.
When performing backward data-flow analysis, the access of fields may break the original flow and hence may lead to unexpected results if not properly handled.

To mitigate this, we transform a field involved in the call trace of the API under testing (in the analyzed app) into a local variable in the generated test case.
The local variable will be initiated following the same method as it is assigned in the original app.
We then search the whole class to check how the field's value is assigned and subsequently apply the same method to initialize the local variable.
If we cannot find the field's assignment or the assigned value is complicated to be reconstructed, we will use a dummy object to mock the required value.

In addition, \tool{} needs to handle branches in the backward dataflow analysis. Specifically, we leverage the Inter-procedural Control Flow Graph ( ICFG~\cite{bodden2012inter}), which is a combination of call-graph(CG) and control flow graph(CFG), to identify the minimum executable context of the target API. Here, CG is a graph representing the calls between methods over the entire program, while CFG is a graph that represents the control flows in a single method. ICFG treats each statement (or a set of sequential statements) as a node, including branch statements that enable path-sensitive analyses, i.e., the propagation of different information along different branches. With ICFG, we are able to implement branch analysis by analyzing the structure of the graph. To this end, for those methods involving multiple branches, \tool{} will separate each branch to form a different test case.

\textbf{I.4 API Parameter Value Inference.}
To support API compatibility testing, e.g., to ensure that the API will, in any case, be reached once the test case is executed, we propose to directly assign possible values to the API's parameters inside the test case, i.e., the test case per se will not contain any parameter.
To achieve this, \tool{} also infers possible values for each parameter of the API to be tested.
This step follows the same strategy i.e., the approach adopted to infer API caller instances, to infer the API's parameter values.  This is done by performing inter-procedural backward data-flow analysis. We apply the same algorithms described in Algorithm~\ref{API_Caller_Instance_Inference} and Algorithm~\ref{Construct_Call_Trace} on each parameter object to figure out the exact value.

Unfortunately, some Android APIs' parameter values may involve sophisticated operations when building their run-time values that are non-trivial to correctly retrieve statically. Specifically, if the analysis process does not end up at a constant/Android system API statement, the value of a parameter is regarded as being undiscovered.
To mitigate this, we introduce a set of pre-defined rules to generate dummy values for such APIs that have their parameter values hard to retrieve practically.
Some of the representative rules for generating dummy values for such hard-to-retrieve parameters are:

\begin{itemize}
\item For the eight primitive data types in Java (such as int, double, etc.) or their wrapper data types (such as Integer, Double, etc.) -- we provide random values for each of them that conform to their types. 
\item For the String  data type in Java -- we generate a random alphanumeric string.
\item For the Array parameter whose basic type is the eight primitive data types (or their wrapper data types) in Java (such as int, Integer, etc.) -- we generate an Array variable with random primitive values.
\item For  Android system-related objects (or the intermediate objects in the calling object's instantiation process), we use a heuristic approach to obtain the corresponding constructors to create their instances. If an object has multiple constructors, we select the simplest one (with the least number of parameters) to achieve the highest possibility of constructing a valid object.

\end{itemize}

\textbf{I.5 Return Value Analysis.}
To support detecting compatibility issues caused by return values (e.g., a given API may return A at API level X and B at API level Y), we propose to output the return value of the target API at the end of the test case. To achieve this, \tool{} adds a statement at the end of the test case to further record the API's return value.

\textbf{I.6 API Usage Construction.}
After obtaining the caller instance, the invocation statements along the call trace and the return object, we can now recover the call sequence from program entry to the target Android API by reversing the retrieved statements step by step.
Based on the results of API caller instance inference along, \tool{} will generate a test case containing the same number and type of parameters as the API to be tested.
For example, as shown in Listing~\ref{code:generated_test_case} at line 3, the generated test case contains five parameters, in the same type and order of the API under testing.

\begin{lstlisting}[
caption={Examples of the generated test cases for API \emph{queryDetailsForUid (int networkType, String subscriberId, long startTime, long endTime, int uid)}.},
label=code:generated_test_case,
firstnumber=1]
//for supporting generic testing
@Test
public void testQueryDetailsForUid(int var1, String var2,long var3, long var4, int var5) throws Exception {
 Context var6 = InstrumentationRegistry.getTargetContext();
 Object var7 = var6.getSystemService("netstats");
 NetworkStatsManager var8 = (NetworkStatsManager) var7;
 var8.queryDetailsForUid(var1, var2, var3, var4, var5);
}

//for supporting compatibility testing
@Test 
public void testQueryDetailsForUid() throws Exception {
 long var1 = System.currentTimeMillis();
 
 Context var2 = InstrumentationRegistry.getTargetContext();
 PackageManager var3 = var2.getPackageManager();
 String var4 = var2.getPackageName();
 ApplicationInfo var5 = var3.getApplicationInfo(var4, 1);
 int var6 = var5.uid;
 
 NetworkStats var7 = testQueryDetailsForUid(0, "", 0L, var1, var6);
// Output return value
 out(var7);
}
\end{lstlisting}

Taking the results of the API parameter value inference step, \tool{} will generate another test case containing no parameters (line 10 in Listing~\ref{code:generated_test_case}).
This test case will directly call the former test case with prepared parameter values.
This test case is specifically designed to support API compatibility testing.
The former test case, on the contrary, is designed to serve a more general purpose.
With the help of fuzzing testing approaches (to generate possible parameter values for the test case), we expect the former test case could be leveraged to discover not only compatibility issues but also design defects such as bugs and security issues.
This trade-off allows \tool{} to generate test cases that are at least suitable for identifying signature-based compatibility issues (e.g., a given API is no longer available in a certain framework), although it may not be effective enough to help identify semantic change involved compatibility issues.

Please note that there are several special classes, such as \emph{InstrumentationRegistry}, involved in the generated unit test cases.
These classes are part of the Android Testing Support Library provided by Google for supporting instrumented unit tests.
Compared to traditional unit tests, also known as local unit tests which can run on the JVM, instrumented unit tests require the Android system to run (e.g., through physical Android devices or emulators).
Since this requires us to generate actual Android apps to run on Android devices or emulators, instrumented tests are much slower than local unit tests.

Nevertheless, we still choose to use instrumented tests to examine the compatibility of Android APIs. This is because instrumented tests provide more fidelity than local unit tests, which we have found is essential to reveal potential compatibility issues, especially those that involve device-specific issues.

\textbf{I.7 Equivalent Test Case Elimination.}
Considering that the constructed test cases could be equivalent (i.e., duplicated), it is necessary to filter them out to save subsequent testing time and resources. Here, based on the concept of semantic equivalence defined in operational semantics~\cite{jiang2009automatic}, we consider that two test cases are equivalent if they share the same API invocation sequence. To this end, we first obtain the API invocation sequence for each test case and then examine the discrepancy between any two of them to check if $O_a$ = $O_b$, where $O_a$ and $O_b$ are the lists of API invocations in two different test cases. Based on this rule, we are able to eliminate equivalent tests (i.e., the first test case is retained). After this step, for the sake of simplicity, if there are still multiple test cases retained for a given API, we will select the small-scale one (with the least number of method invocations) for supporting follow-up analyses.



\vspace{-10pt}

\subsection{App Compatibility Testing}
Using the unit test cases generated by the first module, the second module of \tool{} leverages them to check if the corresponding Android APIs will likely induce app compatibility issues.
It first assembles all the test cases into an Android app and then aggregates their execution results against different devices running different Android frameworks.
We now briefly detail these two steps below.

\textbf{II.1 Test APK Generation.}
As discussed earlier, we have to resort to instrumented unit tests to examine Android APIs' incompatibilities.
This process essentially requires us to generate an Android app (or APK) to be installed and executed on Android systems.
Fortunately, Google has provided such a mechanism to achieve this purpose, i.e., supporting instrumented tests for a limited number of Android APIs. In this work, we directly reuse this mechanism to generate the test APK for all the unit tests automatically generated by \tool{}. 

\textbf{II.2 Test Results Aggregation.}
After the test APK is generated, we can distribute it for execution on multiple devices.
Since the test APK contains only known test cases, it is quite straightforward to execute it fully.
The only challenge that lies in this step is to select the right set of devices on which to execute the test cases to reveal as many incompatible APIs as possible.
Crowdsourced app testing could be an approach to achieve this purpose.

After installing and executing the generated test APK on multiple devices, the last step is in aggregating the test results to highlight potential compatibility issues in the app.
Inspired by the experimental setup of the work proposed by Cai et al.~\cite{cai2019large}, we consider an API as a potential incompatible case if 1) its corresponding test case can successfully run on a nonempty set of devices while failing on others; 2) its corresponding test case returns different values when running on different SDKs.

\section{Evaluation}
%
%
Our \tool{} aims to generate unit test cases covering as many Android APIs as possible, so as to allow the discovery of more API-induced compatibility issues in apps.
To evaluate if this goal has been fulfilled, we propose to answer the following three research questions.

\begin{description}
\setlength\itemsep{0.05em}

\item[RQ1] To what extent can \tool{} generate executable unit test cases for Android APIs?


\item[RQ2] How effective is \tool{} in discovering API-induced Compatibility Issues? 

\item[RQ3] How does \tool{} compare with existing tools in detecting compatibility issues?


\end{description}


\subsection{Experimental Setup}
To investigate the success rate of \tool{} in producing valid test cases, we randomly select 1,000 Android apps for each target SDK version between 21 (i.e., Android 5.0) and 30 (i.e., Android 11.0\footnote{The latest version at the time when we conducted this study.}) from AndroZoo to prepare the experimental dataset. 
Here, we select 1,000 apps for each target SDK version because compatibility issues mainly lie in the evolution of APIs on different Android SDK versions~\cite{li2018cid}. Here, the criteria for app selection are based on the targetSdkVersion, which is the most appropriate API level on which the app is designed to run. Hence, the overall dataset for the experiment contains 10,000 Android apps whose target API versions are distributed equally across ten API levels. 
Our experiment runs on a Linux server with Intel(R) Core(TM)
i9-9920X CPU @ 3.50GHz and 128GB RAM. The timeout setting for analyzing each app with \tool{} is 20 minutes. 
In this experiment, we generate the test cases for all Android APIs that have been invoked in the apps. However, users of \tool{} could provide a customized list of APIs to only generate test cases based on their interests.

\subsection{RQ1-Effectiveness}

Our first research question concerns the effectiveness of \tool{} in mining API usages from existing Android apps to generate valid unit tests for Android APIs. In this work, we consider a test case to be valid if (1) the generated code snippet can be successfully compiled on all API levels and (2) the test case does not throw an exception before the execution point of the API on all API levels.
The first condition ensures that the test case is syntactically correct. The second condition makes sure that the API's execution environment is properly set up. In other words, the second condition ensures that the exceptions we collected from valid test cases are exceptions thrown by the API under testing, which is essential for examining if the API will induce compatibility issues.

\begin{figure}[t!]
    \centering
    \setlength{\belowcaptionskip}{1pt}
    \includegraphics[width=0.35\textwidth]{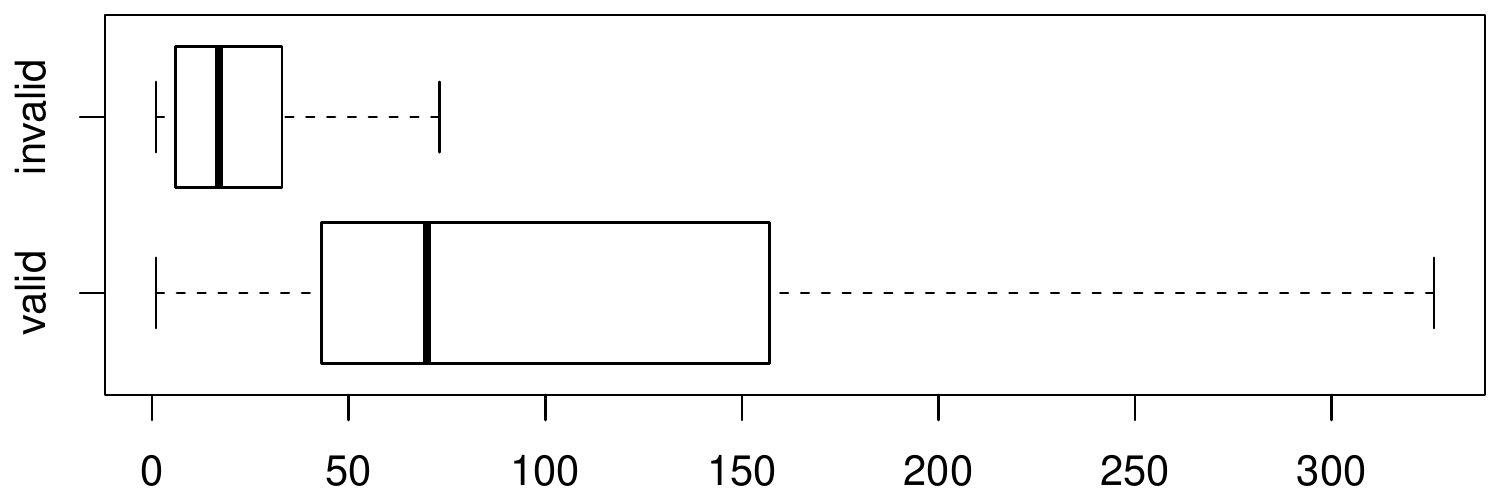}
	\caption{Distribution of the number of valid and invalid test cases per APK.}
    \label{fig:valid_per_apk}
\end{figure}

\textbf{Result.}
Among the 10,000 randomly selected apps, \tool{} generates in total 1,032,182 test cases.
After eliminating the equivalent test cases, 66,499 of them are retained as distinct test cases, w.r.t. 28,367 distinct Android APIs. For the sake of simplicity, we select the small-scale test case (with the least number of invocation sequences) for each API (i.e., 28,367 test cases) for further study. By compiling and executing these 28,367 test cases, we further confirm that 5,562 of them are invalid (i.e., 22,805 of them are valid), giving a success rate of 80.4\% in generating valid test cases. In addition, our manual analysis on 100\footnote{The sample size is determined based on a confidence level at 95\% and a confidence interval at 10(\url{https://www.surveysystem.com/sscalc.htm}).} randomly selected test cases confirm that these test cases generated by \tool{} are indeed valid.


Figure~\ref{fig:valid_per_apk} summarizes the distribution of the number of valid and invalid test cases generated per app.
The median number of valid and invalid test cases generated per app are 70 and 17, respectively, while their average are 106.29 and 25.37, respectively. Like most other state-of-the-art approaches, even though our static analysis approach has limitations so that it cannot generate valid test cases for every API, especially the complex ones, \textbf{our approach is still capable of generating more valid test cases than invalid ones.}
This demonstrates the effectiveness of our approach in mining Android API usages to generate API unit test cases. 



We note that the success rate of generating valid test cases  -- ~80.4\% at the moment -- is important but not crucial to our work. Theoretically, as long as we increase the number of Android apps considered for learning, we would likely be able to generate valid test cases for the given API under testing. For the test cases that are regarded as invalid, we further manually look into their root causes.
Our in-depth analysis reveals that the invalid cases are mainly caused by the lack of prerequisites (e.g., resource files), especially in UI-related APIs. For example, UI objects can hardly be programmatically initialized without certain resource files.



\begin{figure}[t!]
    \centering
    \setlength{\abovecaptionskip}{1pt}
    \includegraphics[width=0.3\textwidth]{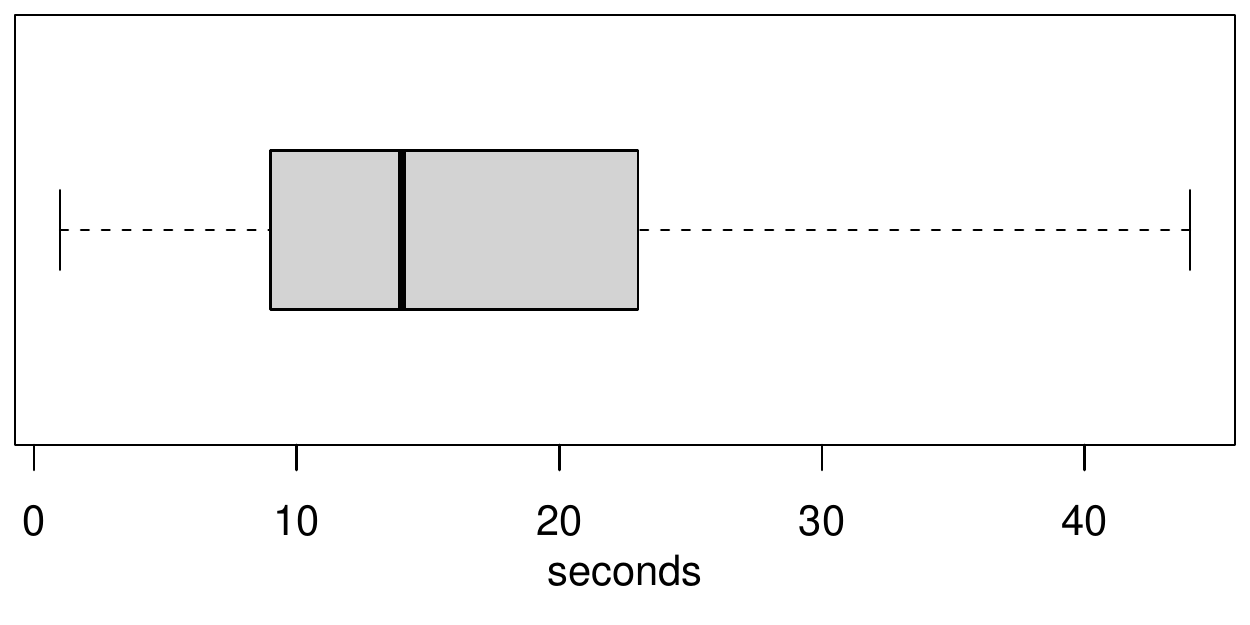}
	\caption{Distribution of time spent by \tool{} to generate test cases per APK.}
    \label{fig:time_cost}
\end{figure}

We further look at the efficiency of \tool{} by reporting its time cost when applied to generate test cases. Figure~\ref{fig:time_cost} summarizes the distribution of time consumed by \tool{} per app. On average, it takes 17.45 seconds to process an app. The median time cost is 14 seconds. The fact that the time spent by \tool{} to process an Android app is quite small suggests that it is practical to apply \tool{} to analyze large-scale Android apps.

\begin{tcolorbox}[title=\textbf{Answers to RQ1}, left=2pt, right=2pt,top=2pt,bottom=2pt]
By learning from existing API usages, our approach can automatically generate API unit test cases. Despite various challenges posed by advanced Android programming features, our static analysis approach can still achieve over 80\% of the success rate in generating valid test cases.
\end{tcolorbox}

\subsection{RQ2-Performance on Real-World Applications}
The ultimate goal of \tool{} is to help better identify API compatibility issues that occur during Android system evolution. To this end, in this research question, we evaluate, based on the generated valid unit test cases, to what extent can \tool{} help in identifying API-induced compatibility issues.

We consider an Android API has a compatibility issue if the execution results on different Android SDK versions are inconsistent. More specifically, an API is considered to have a compatibility issue if any of the following happens: (1) the corresponding test case runs successfully on some Android SDK versions but fails (e.g., throws errors or exceptions) on others; or (2) the corresponding test case throws different errors or exceptions on different Android SDK versions (e.g., throws \emph{NoSuchMethodError} on some versions, while throws \emph{IllegalArgumentException} on others); or (3) the return values of a target API are non-identical on different SDK versions.



\textbf{Result.}
Recall that \tool{} successfully generates 22,805 distinct valid test cases covering 22,805 unique Android APIs, based on the randomly selected 10,000 apps. In this work, these 22,805 test cases are respectively executed on ten Android emulators running API levels from 21 to 30.

By comparing the experimental results against the aforementioned three rules, \textbf{we are able to locate 3,488 compatibility issues covering 2,695 Android APIs.} Note that some APIs may involve more than one compatibility issue.
To confirm whether the APIs identified by \tool{} indeed have compatibility issues, we manually examined 100 randomly selected APIs reported to have compatibility issues, 100 of them are confirmed to have compatibility issues (i.e., true positive results). Here, we remind the readers that all of the compatibility issues are actually identified through dynamic analysis, which is expected to be highly accurate. In the manual validation process, we manually examine the APIs' implementation in Android framework source code across different SDK versions and compare it with the release notes in the official API documentation.

According to the root causes of the compatibility issues, we further categorize them into the following types. 
\begin{itemize}
\item \textbf{Type 1: Signature-based compatibility issue.} This type refers to the incompatibility caused by adding new APIs, deprecating existing APIs, or changing existing APIs' signatures, such as changing parameters or return types.
    
\item \textbf{Type 2: Semantics-based compatibility issue.} This type refers to the incompatibility caused by the same API (i.e., the signature is not changed) behaving differently on different Android API levels. Based on the APIs' behaviours, semantic-based compatibility issues are further breakdown into the following two sub-types.

\textbf{1) Type 2.1 Semantics-based compatibility issue: Different Errors}: The APIs categorized into this type will throw exceptions or errors (including app crashes) on devices running some SDK versions but will not (either running successfully or throw different exceptions/errors) when running on devices with other SDK versions. \par
\textbf{2) Type 2.2 Semantics-based compatibility issue: Different Return Values}: APIs of this type will not directly cause compatibility issues. Under the same experimental setting, these APIs will return different values when running on different SDKs.
However, if the different return values are not specifically distinguished, the subsequent code that uses these return values may behave differently on different devices, leading to also compatibility issues.
\end{itemize}

Among the 3,488 identified compatibility issues, 438 of them suffer from signature-based issues, while 3,050 are caused by semantic issues. Within the semantic-based compatibility issues, 946 of them are caused by different errors (Type 2.1), while the remaining 2,104 cases observe different return values (Type 2.2). Table~\ref{tab:catgory_exception} summarizes the possible errors/exceptions that cause the signature/semantic compatibility issues.

As expected, the most common error is no such method error, which can be caused by (1) the API being deprecated and removed from the framework, or (2) the API not yet introduced.
Both reasons are introduced by the evolution of the framework.
As also revealed by Li et al.~\cite{li2018cid}, the fast evolution of the Android framework has indeed introduced a lot of APIs that will likely induce compatibility issues.
Except for signature-based issues, which are relatively easy to be statically identified (for example, by comparing the framework codebase of different versions), our approach has also found 2,974 issues (over five times the number of Type 1 issues) involving semantic changes of APIs, which are non-trivial to be identified statically~\cite{liu2021identifying}.

\begin{table}[!h]
\scriptsize
\centering
\setlength{\belowcaptionskip}{1pt}
\caption{Categories and statistics of the observed error/exception types associated with compatibility issues.} 
\label{tab:catgory_exception}
\resizebox{0.85\linewidth}{!}{
\begin{tabular}{l l c} 
\hline
Issue Type &  Errors/Exceptions & Count\\
 \hline
\multirow{3}{*}{Signature}       & 
 NoSuchMethodError   & 270 \\
 &  NoClassDefFoundError & 163  \\
& NoSuchFieldError & 5\\

\\
\multirow{27}{*}{Semantic}  & 
 SecurityException  & 196\\
 &  NullPointerException & 189\\
 &  RuntimeException & 139\\
 & Resources\$NotFoundException & 113 \\
 &  IllegalArgumentException & 67 \\ 
 &  NoSuchElementException & 42\\
 & Crash & 41 \\
 &  IllegalStateException & 24\\
 & AndroidRuntimeException & 23\\
 & IOException & 15 \\
 &  ArrayIndexOutOfBoundsException & 15\\
 & FileNotFoundException & 12\\
 & PackageManager\$NameNotFoundException & 10 \\
 & IndexOutOfBoundsException & 10 \\
 & ClassCastException & 9 \\
 & IllegalAccessError & 9 \\
 & ActivityNotFoundException & 8 \\
 & StringIndexOutOfBoundsException & 5 \\
 & UnsupportedOperationException & 5 \\
 & BadTokenException & 3 \\
 & CanceledException & 3 \\
 & ErrnoException & 2 \\
 & KeyChainException & 2 \\
 & SQLiteCantOpenDatabaseException & 1 \\
 & ParseException & 1 \\
 & StackOverflowError & 1 \\
 & NumberFormatException & 1 \\
\hline
\end{tabular} 
}
\end{table}

Below, we elaborate on real-world compatibility issues for each type of case study.

\textbf{Case Study 1: Signature-based Compatibility Issue.}
The API \emph{android.content.pm.LauncherApps\#hasShortcutHostPermission} has been reported to contain a signature compatibility issue. The corresponding test case\emph{(whose API usage is extracted from app ch.deletescape.\\lawnchair.plah\footnote{SHA-256:bf4e6e7fb594cd9db4b168a68f70157ad9c1fea0192e0bd5d9a39d1c38802639})} throws \emph{NoSuchMethodError} on Android SDK version 21 to 23 but can be successfully executed on Android SDK version 24 to 30. This result suggests that the API was introduced to the Android system since API level 24; therefore, it would cause an error if the containing app runs on devices with Android SDK version earlier than 24. 
However, according to the official Android API documentation, this API was added in API level 25 \cite{hasShortcutHostPermission}, which is imprecise according to our result. We further checked the source code of Android SDK 24 and confirmed its existence. 


\textbf{Case Study 2: Semantics-based compatibility issue caused by different Exceptions.}
The API \emph{android.app.NotificationManager
\#notify} has been reported to contain a semantic compatibility issue. The corresponding test case\emph{(whose API usage is extracted from app com.ag.dropit\footnote{SHA-256:30f7f72cebeffd7c6e26489198ee5ad244bd44b076dd9cb59865d8b0e82a86af})} can be successfully executed on Android SDK version 21 to 22 but throws \emph{IllegalArgumentException} on Android SDK version 23 to 30. We manually looked into its source code in the Android codebase and found that the actual implementation of this API has been changed since API level 23, which added a sanity check of object \emph{mSmallIcon}. This explains why it throws \emph{IllegalArgumentException} when there is no valid small icon from the API level after 23. Unfortunately, the official Android API documentation does not reflect this change, which is misleading.

\textbf{Case Study 3: Semantics-based compatibility issue caused by different return values.}
The API \emph{(extracted from app cleaner.\\clean.booster\footnote{SHA-256:def5db37b3a68de62a0472e872700092060bdec3e875d4f476fcda52795bceb2})} \emph{<android.text.format.Formatter: String formatShortFileSize(Context, long)>} has a return value-induced compatibility issue. The format of the return values vary on different API levels: given the 1L(The long data type of value 1) as the second parameter, the return value on API level 21 to 22 is \emph{``1.0B''}, the return value on API level 23 is \emph{``1.0 B''} (with additional whitespace between 1.0 and B), while the return value on API level 24 to 30 is \emph{``1 B''}. The difference in return values can introduce potential issues if app developers rely on the return value to implement future functions without checking the running API level.
For example, if app developers cast the return value from String to Byte afterwards, it may throw an exception if users ignore the value discrepancy on different API levels.

\begin{tcolorbox}[title=\textbf{Answers to RQ2}, left=2pt, right=2pt,top=2pt,bottom=2pt]
Our approach is useful in automatically pinpointing API-induced compatibility issues. It also goes beyond the state-of-the-art to be capable of detecting not only signature-based compatibility issues but also more significant semantics-based compatibility issues, i.e., the corresponding APIs' signatures are kept the same, while their semantics are altered.
\end{tcolorbox}

\subsection{RQ3-Comparison with State-of-the-art}
Considering the main purpose of our work is generating test cases for detecting compatibility issues, both generic test case generation approaches, such as EvoSuite~\cite{fraser2011evosuite} and compatibility issues detection tools, such as CiD~\cite{li2018cid}, are selected as the baselines to evaluate our approach. We evaluate the performance of \tool{}, EvoSuite~\cite{fraser2011evosuite} and CiD~\cite{li2018cid} in detecting compatibility issues. Overall, table~\ref{tab:comparison_RQ3} lists the number of compatibility issues found by \tool{}, EvoSuite and CiD. We then break down the comparative results as follows:

\begin{table}[t!]
\centering
\caption{The comparison results between {\tool{}} and Evosuite, CiD.} 
\label{tab:comparison_RQ3}
\resizebox{0.7\linewidth}{!}{
\begin{tabular}{r | c c c} 
\hline
Tool & \# Type 1 & \# Type 2.1 & \# Type 2.2 \\
\hline
{\tool{}} & 438  & 946 & 2,104  \\
Evosuite & 36 & 0  & 0  \\
CiD & 864  & -  & -  \\
\hline
\end{tabular} }
\end{table}

\textbf{Comparison with EvoSuite.}
To compare \tool{} with generic test case generation tools, we choose EvoSuite as the baseline because EvoSuite has been considered the state-of-the-practice test generation tool, which achieved the highest score at the SBST 2021 Tool Competition~\cite{vogl2021evosuite}. EvoSuite uses an evolutionary search approach to generate and optimize test suites toward satisfying an entire coverage criterion for Java classes. We remind the readers that the objective of EvoSuite is to generate tests for classes, not directly aiming at generating tests for APIs, which are the main target when concerning compatibility issues happening in Android apps.
Since Evosuite can only generate tests based on source code, we resort to AOSP from SDK 21 to 30 for Evosuite to generate test cases.
In total, Evosuite successfully generates 5,335 test cases. 
We then execute all of them on SDK versions from 21 to 30 and apply the same rules for determining compatibility issues as used in \tool{}. 

As shown in Table~\ref{tab:comparison_RQ3}, EvoSuite only finds 36 signature-based compatibility issues, and no semantic compatibility issues are identified. We further manually check the test cases generated by EvoSuite and observe that the false negatives (compared with \tool{}) are mainly caused by overlooking API dependency information. For example, some APIs can only be invoked by system services, which makes EvoSuite fail to generate valid tests without knowing the usage of such APIs. Missing API dependency information makes EvoSuite insufficient in pinpointing compatibility issues. In other words, our comparison result reveals that mining API usage from apps is beneficial for finding compatibility issues.

\textbf{Comparison with CiD.}
To the best of our knowledge, no work has been devoted to detecting compatibility issues in Android apps dynamically. CiD~\cite{li2018cid} is the closest work to ours on detecting compatibility issues. CiD model and compare API signatures on different SDK versions to detect compatibility issues. We thus evaluate the performance of \tool{} and CiD on the same dataset in RQ1, which contains 10,000 Android apps. 

In total, CiD detects 864 compatibility issues, as highlighted in Table~\ref{tab:comparison_RQ3}.
Out of the 864 compatibility issues detected by CiD, \tool{} successfully identified 3,050 cases that have been overlooked by CiD. We further randomly analyze 50 false negatives of CiD and find that these issues are caused by the lack of semantics analysis of the API implementation. Specifically, when analyzing the evolution of APIs, CiD only examines the change of API signatures (including name, type, and parameters); hence it is not capable of detecting APIs that modify the implementation details but retain the same signature. Also, we find other 51 false positives of CiD are caused by imprecisely extracting the usage of the APIs, due to the context-insensitive approach of CiD when building the conditional call graph (CCG).
On the other hand, we find that \tool{} miss 375 cases reported in CiD (i.e., false negatives in \tool{}). This is mainly caused by the sophisticated usage of some APIs, which cannot easily be initialized programmatically (e.g. UI-related APIs). For example, some APIs may involve the initialization of UI objects that cannot be initialized programmatically. This makes it very challenging to automatically generate unit test cases in some circumstances.
However, we argue that this limitation can be alleviated by manually adding prerequisite resources to create more valid tests.
Overall, the comparison results reveal the weakness of static analysis approaches in detecting semantic compatibility issues, i.e., yields false-positive results and is hard to detect issues involving semantic changes in methods. 
It also demonstrates that our approach can indeed find more diverse compatibility issues and hence is promising to complement existing static approaches.



\begin{tcolorbox}[title=\textbf{Answers to RQ3}, left=2pt, right=2pt,top=2pt,bottom=2pt]
\tool{} outperforms the state-of-the-practice test generation tool, EvoSuite, and the state-of-the-art static compatibility detection tool, CiD, in pinpointing compatibility issues caused by the fast evolution of Android APIs. 
This experimental evidence shows the necessity to perform dynamic testing to pinpoint compatibility issues in Android apps, and it should take API usage dependencies into consideration when generating test cases to fulfill the dynamic testing approach.
\end{tcolorbox}



\vspace{-3pt}
\section{Discussion}

We now discuss the potential implications and limitations of this work.

\subsection{Implications}
\textbf{Better Supplementing Compatibility Analysis: } 
\tool{} is able to automatically generate tests for Android APIs for detecting both signature-based and semantics-based compatibility issues. Previous static analysis works~\cite{li2018cid} overlooked the semantics-based ones (i.e., APIs have the same signature but different implementation), which are more challenging to detect statically. Together with the state-of-the-art static analysis-based methods, our proposed method can provide a more comprehensive overview of compatibility issues in Android APIs. 
Therefore, we argue that there is a need to invent hybrid approaches to take advantage of both static analysis and dynamic analysis to conquer compatibility issues.

\textbf{Beyond Compatibility Testing: }
\tool{} is not only useful in pinpointing compatibility issues in Android APIs but also could be easily adopted to automatically generate test cases for other purposes.
For instance, in the cases that an API takes various parameters as input, it can work with other testing approaches such as fuzz testing to explore the API's implementation dynamically.
It hence goes beyond compatibility testing and provides a more general-purpose form of API testing.

\textbf{Go Beyond Android.}
Our approach performs static program analysis to learn and generate test cases from existing API usages, which are not strongly attached to Android apps. We believe it could be easily adapted to analyze other Java projects, e.g., to automatically generate test cases for popular Java libraries. Although our approach cannot be directly applied to analyze projects written in other programming languages than Java, we believe the idea and methodology proposed in this paper could still work.
We plan to explore these research directions in our future work. We also encourage our fellow researchers to explore this direction further.

\subsection{Limitations}
The main limitation of \tool{} lies in its backward data-flow analysis when inferring API caller instance and API parameter values.
Indeed, as already known in the community, it is non-trivial to implement a sound data-flow analysis.
Other researchers often accept trade-offs to obtain relatively good results, and this is the same in our case.
When the variables to be backwardly retrieved are complex, e.g., involving constructing various intermediate objects, \tool{} rely on their simplest constructor to initialize the objects to generate a valid test case. Unfortunately, some of the constructors are too complex to initialize, leading to invalid test cases. Some other APIs may involve the initialization of UI objects that cannot be initialized programmatically (hence random values are leveraged to handle such cases). This makes it very challenging to automatically generate unit test cases in all circumstances.

Second, currently, our \tool{} data-flow analysis is agnostic to some advanced programming features, such as reflective and native calls. This may further impact the soundness of our approach.
As part of our future work, we plan to integrate approaches developed by our fellow researchers to mitigate those long-standing challenges (e.g., applying DroidRA~\cite{li2016droidra} to mitigate the impact of reflective calls on our static analysis approach.).

Third, the capability of our approach is limited by the number of Android APIs leveraged by real-world apps. Our approach cannot generate unit test cases for such APIs that have never been accessed by real-world Android apps.
Nevertheless, we argue that this impact is not significant as the APIs that have never been used by app developers should have a low priority to be tested than the others that are frequently accessed.
Compared to the latter case, the former ones will not cause problems such as crashes to Android apps.
Subsequently, it will not impact the user experience and the reputation of the app developers.

Fourth, considering the generated parameter values are inferred from real-world apps, which might not reveal all possible semantic-related compatibility issues. In other words, the capability of our approach is limited by the values of parameters leveraged by real-world apps. As for our future work, we plan to integrate fuzzing techniques~\cite{zhang2020bigfuzz,patra2016learning, sui2011effective} into our approach so as to trigger as many compatibility issues as possible.

Fifth, our definition of compatibility issue is based on the observation that the same test case throws different errors or exceptions on different Android SDK versions. However, this might introduce false negatives because the tests that throw the same exception across all SDK versions are ignored. Nevertheless, we argue that this type of false-negative requires further human validation and thus cannot be determined automatically. In addition, related works~\cite{cai2019large} also have not considered this situation.

Sixth, the tests generated by our approach may suffer from flaky tests. Indeed, non-deterministic return values may appear under different execution environments, leading to false positives. However, it is a non-trivial task to tackle flaky tests ~\cite{zolfaghari2021root} because the root causes of flaky tests can be quite sophisticated. Nevertheless, as part of our future work, we plan to integrate other approaches developed by our fellow researchers to mitigate this long-standing challenge, e.g., by applying FlakeScanner~\cite{dong2021flaky} to mitigate the impact of flaky tests on our approach.

Seventh, the types of compatibility issues detected in our approach are the ones that are related to exceptions or return values. However, this will certainly not be complete to cover all possible cases. For example, the value of variables in the same API may evolve on different SDK versions. Indeed, as summarised by Liu et al.~\cite{liu2022automatically}, apart from compatibility issues raised by API signature/semantic changes and return value differences, there exist other types of compatibility issues, such as those introduced by field evolution, callback method changes, etc. Nevertheless, as also highlighted by Liu et al.~\cite{liu2022automatically}, the number of such compatibility issues is quite limited, suggesting that the impact of such cases on our approach may not be significant.

Eighth, the main objective of the generated test cases in this work is to pinpoint compatibility issues.
The quality of these test cases (e.g., readability, overlaps, and maintenance) has hence not been considered. We therefore commit to further investigating the quality of the generated test cases in our future work.

Last but not least, our approach relies on existing code examples to generate test cases. However, the selected code examples may contain sub-optimal or erroneous API usages. Nevertheless, we argue that this impact is not significant as we extract code examples from real-world Android apps from Google Play, for which thousands of users might have used (hence tested) them in practice.

\section{RELATED WORK}

Android API evolution is a critical issue in software maintenance ~\cite{nagappan2016future,martin2016survey,li2016accessing,oliveira2018android,lamothe2020a3,li2018characterising,zhou2016api,dig2005role,kapur2010refactoring,sun2021characterizing}. McDonnell et al. ~\cite{mcdonnell2013empirical} have shown that the Android
system updates 115 APIs per month on average, while app developers usually adopt the new APIs much more slower. The slow adoption of API updates may raise various issues, such as security and compatibility. An empirical study on StackOverflow conducted by Linares-Vasquez et al. ~\cite{linares2014api} suggests that API updates would trigger more discussions, especially if APIs are removed from the Android system. They also revealed that users are in more favour of apps that use less fault and change-prone APIs \cite{linares2013api, bavota2014impact}, as these apps would likely introduce fewer failures, crashes and other bugs.


Android developers have long been suffered from compatibility issues due to the fast-evolving and fragmented nature of the Android ecosystem \cite{xia2020android,kamran2016android,li2017static,nayebi2012state}. 
Researchers have proposed several solutions for detecting compatibility issues of Android APIs. Wei et al. \cite{wei2016taming, wei2018understanding} conducted an empirical study to investigate fragmentation-induced API compatibility issues and proposed a tool named FicFinder to detect such APIs. FicFinder identifies APIs with compatibility issues based on heuristic rules manually derived from a limited number of Android apps, which is expected to introduce high false negatives (i.e., missing undiscovered compatibility issues). Thereafter, several works have been proposed to leverage data-driven techniques that automatically mine compatibility issues from various sources such as Android code base and real-world apps. Li et al. proposed a tool named CiD to detect potential compatibility issues by mining the history of the Android framework source code. CiD identifies Android APIs' lifetimes and finds if an app's declared supported versions conflict with its used APIs. 

Comparable to our method, several works have been proposed to mine API usage from real-world apps. Scalabrino et al. \cite{scalabrino2019data, scalabrino2020api} considered the APIs wrapped in a version check condition (e.g., \emph{if (Build.VERSION.SDK\_INT >= 21)}) to potentially have compatibility issues and developed a tool named ACRYL to extract such APIs from real-world apps. However, ACRYL can only detect APIs whose compatibility issue is already known by the developers (i.e., they are enclosed in the version check conditions by the developers), while our method is capable of detecting zero-day compatibility issues that the app developers are not yet aware of, or even Google itself. 
Other generic test generation tools, such as EvoSuite and Randoop\cite{pacheco2007randoop}, are able to generate tests for Java classes. However, these tools do not directly aim at generating tests for APIs and they have been demonstrated as insufficient in pinpointing compatibility issues because of the lack of API usage knowledge. 

\section{CONCLUSION}
In this work, we presented a novel prototype tool, \tool{}, that mines existing Android API usages to generate API-focused unit test cases automatically for pinpointing potential compatibility issues caused by the fast evolution of the Android framework.
Experimental results on thousands of real-world Android apps show that (1) \tool{} is capable of automatically generating valid unit test cases for Android APIs with an 80.4\% success rate;
(2) the automatically generated test cases are useful for pinpointing API-induced compatibility issues, including not only signature-based but also semantics-based compatibility issues; and (3) \tool{} outperforms the state-of-the-practice test generation tool, EvoSuite, and the state-of-the-art static compatibility detection tool, CiD, in pinpointing compatibility issues.

\section*{Acknowledgements}
The authors would like to thank the anonymous reviewers who have provided insightful and constructive comments on this paper. This work was partly supported by the Australian Research Council (ARC) under a Laureate Fellowship project FL190100035, a Discovery Early Career Researcher Award (DECRA) project DE200100016, and a Discovery project DP20010002.

\balance
\bibliographystyle{acm}
\bibliography{ref}

\end{document}

%% file: predefine.tex
\usepackage{multirow}    					
\usepackage{hhline}     				 		
\usepackage{algorithm} 
\usepackage{algpseudocode}
\usepackage{enumerate}
\usepackage{wrapfig}
\usepackage{fancybox}
\usepackage{url}
\usepackage{balance}
\usepackage{pifont}
\usepackage{xcolor}
\usepackage{tikz}
\usepackage{graphicx}
\usepackage{color}    
\usepackage{listings}
\usepackage{siunitx}
\usepackage{soul} 
\usepackage[inline]{enumitem} 

\definecolor{KWColor}{rgb}{0.37,0.08,0.25}
\definecolor{CommentColor}{rgb}{0.133,0.545,0.133}
\definecolor{StringColor}{rgb}{0,0.126,0.941}